\documentstyle[12pt]{article}
\input psfig.sty

\def\MPL #1 #2 #3 {Mod.~Phys.~Lett.~{\bf#1},\  #2 (#3)}
\def\NPB #1 #2 #3 {Nucl.~Phys.~{\bf#1},\  #2 (#3)}
\def\PLB #1 #2 #3 {Phys.~Lett.~{\bf#1},\  #2 (#3)}
\def\PR #1 #2 #3 {Phys.~Rep.~{\bf#1},\ #2 (#3)}
\def\PRD #1 #2 #3 {Phys.~Rev.~{\bf#1},\  #2 (#3)}
\def\PRL #1 #2 #3 {Phys.~Rev.~Lett.~{\bf#1},\  #2 (#3)}
\def\RMP #1 #2 #3 {Rev.~Mod.~Phys.~{\bf#1},\  #2 (#3)}
\def\ZP #1 #2 #3 {Z.~Phys.~{\bf#1},\  #2 (#3)}
\def\IJMP #1 #2 #3 {Int.~J.~Mod.~Phys.~{\bf#1},\  #2 (#3)}

\def\ebtag{e_{b-tag}}
\def\ectag{e_{c-tag}}
\def\emistag{e_{mis-tag}}
\def\h{h}

\def\rts{\sqrt s}
\def\ie{{\it i.e.}}

\def\anti{\overline}

\def\h{h}

\def\tanb{\tan\beta}
\def\hl{h^0}

\def\ha{A^0}
\def\mha{m_{\ha}}
\def\hh{H^0}

\def\fbi{~{\rm fb}^{-1}}

\def\gev{~{\rm GeV}}
\def\tev{~{\rm TeV}}
\def\mstop{m_{\widetilde t}}
\def\mt{m_t}

\def\overlay#1#2{\ifmmode \setbox 0=\hbox {$#1$}\setbox 1=\hbox to\wd 0{\hss
$#2$\hss }\else \setbox 0=\hbox {#1}\setbox 1=\hbox to\wd 0{\hss #2\hss }\fi
#1\hskip -\wd 0\box 1}
\def\case#1/#2{{\textstyle{#1\over#2}}}
\def\9{\phantom 0}      
\renewcommand\linebreak{\unskip\break} 
\def\lsim{\mathrel{\raise.3ex\hbox{$<$\kern-.75em\lower1ex\hbox{$\sim$}}}}
\def\gsim{\mathrel{\raise.3ex\hbox{$>$\kern-.75em\lower1ex\hbox{$\sim$}}}}
\catcode`@=11
\newcount\@tempcntc
\def\@citex[#1]#2{\if@filesw\immediate\write\@auxout{\string\citation{#2}}\fi
  \@tempcnta\z@\@tempcntb\m@ne\def\@citea{}\@cite{\@for\@citeb:=#2\do
    {\@ifundefined
       {b@\@citeb}{\@citeo\@tempcntb\m@ne\@citea\def\@citea{,}{\bf ?}\@warning
       {Citation `\@citeb' on page \thepage \space undefined}}%
    {\setbox\z@\hbox{\global\@tempcntc0\csname b@\@citeb\endcsname\relax}%
     \ifnum\@tempcntc=\z@ \@citeo\@tempcntb\m@ne
       \@citea\def\@citea{,}\hbox{\csname b@\@citeb\endcsname}%
     \else
      \advance\@tempcntb\@ne
      \ifnum\@tempcntb=\@tempcntc
      \else\advance\@tempcntb\m@ne\@citeo
      \@tempcnta\@tempcntc\@tempcntb\@tempcntc\fi\fi}}\@citeo}{#1}}
\def\@citeo{\ifnum\@tempcnta>\@tempcntb\else\@citea\def\@citea{,}%
  \ifnum\@tempcnta=\@tempcntb\the\@tempcnta\else
   {\advance\@tempcnta\@ne\ifnum\@tempcnta=\@tempcntb \else \def\@citea{--}\fi
    \advance\@tempcnta\m@ne\the\@tempcnta\@citea\the\@tempcntb}\fi\fi}
\catcode`@=12

\renewenvironment{thebibliography}[1]
 {\begin{list}{\arabic{enumi}.}
    {\usecounter{enumi} \setlength{\parsep}{0pt}
     \setlength{\itemsep}{3pt} \settowidth{\labelwidth}{#1.}
     \sloppy
    }}{\end{list}}

\def\rta{\rightarrow}
\def\tanb{\tan\beta}

\begin{document}

\newlength{\captsize} \let\captsize=\small 

%
\font\fortssbx=cmssbx10 scaled \magstep2
\hbox to \hsize{
$\vcenter{
\hbox{\fortssbx University of California - Davis}
}$
\hfill
$\vcenter{
\hbox{\bf UCD-96-21} 
\hbox{July, 1996}
}$
}

\medskip
\begin{center}
\bf
Detection of Neutral MSSM Higgs Bosons in Four-{\boldmath $b$}
Final States at the Tevatron and the LHC: An Update
\\
\rm
\vskip1pc
{\bf J. Dai$^a$, J.F. Gunion$^a$, and R. Vega$^c$}\\
\medskip
\small\it
$^a$Physics Department, University of California, La Jolla, CA 92093, USA\\
$^b$Davis Institute for High Energy Physics, 
University of California,  Davis, CA 95616, USA\\
$^c$Physics Department, Southern Methodist University, Dallas, TX 75275, USA\\
\end{center}

\begin{abstract}

We update our earlier results regarding
detection of $gg\rta b\anti b \h\rta 4b$ ($\h=\hl,\hh,\ha$)
to incorporate the very high
$b$-tagging efficiencies and purities that are now anticipated
at the LHC.  New results for the Tev$^*$ are given,
and indicate substantial potential for these modes.
The complementarity of the $\hh\rta\hl\hl\rta 4b$ final
state mode is illustrated for the LHC. The latest
radiative corrections to the Higgs sector are incorporated.

\end{abstract}

\bigskip

In an earlier paper \cite{DGV3}
we explored the possibility of finding one or more of
the neutral Higgs bosons
of the minimal supersymmetric standard model (MSSM) 
in $gg\rta b\anti b\h$ ($\h=\hl,\hh,\ha$) production followed
by $\h\rta b\anti b$. Vastly superior
$b$-tagging efficiency and purity is now deemed feasible at the LHC,
relative to the conservative assumptions of the earlier work.
Canonical values now employed by ATLAS and CMS \cite{fgianotti} are: 
$\ebtag\sim 0.6$ and $\emistag\sim 0.01$
for $|\eta|\leq 2.5$ and $p_T\geq 15\gev$ at low luminosity (applicable
for accumulated luminosity of $L=30\fbi$ per detector); and $\ebtag\sim 0.5$
and $\emistag\sim0.02$ for $|\eta|\leq 2.5$ and $p_T\geq 30\gev$
at high luminosity ($L=300\fbi$ per detector accumulated).\footnote{The
quantities $\ebtag$ and $\emistag$ are, respectively,
the probabilities for tagging a real $b$-jet and of mis-tagging
a light-quark or gluon jet as a $b$-jet.}
Further, excellent $b$-tagging capabilities are now anticipated
at the Tevatron upgrades, especially the Tev$^*$.
At an upgraded Tevatron with upgraded
detectors, CDF and D0 now expect to achieve $\ebtag\sim 0.5$ with $\emistag\sim
0.005$ for jets with $p_T\gsim 15\gev$ and $|\eta|\leq2$,
at instantaneous luminosities capable of yielding $L=10-30\fbi$ per 
year \cite{womersley}. At both the LHC and Tevatron detectors,
it is now estimated that the probability for tagging a $c$-quark
jet as a $b$-quark jet is $\ectag\sim\ebtag/3$.
In this note, we update our earlier LHC results to incorporate the
improved $b$-tagging expectations, and give, in addition,
new results for these same modes at the Tev$^*$.

\begin{figure}[[htbp]
\vskip 1in
\let\normalsize=\captsize   
\begin{center}
\centerline{\psfig{file=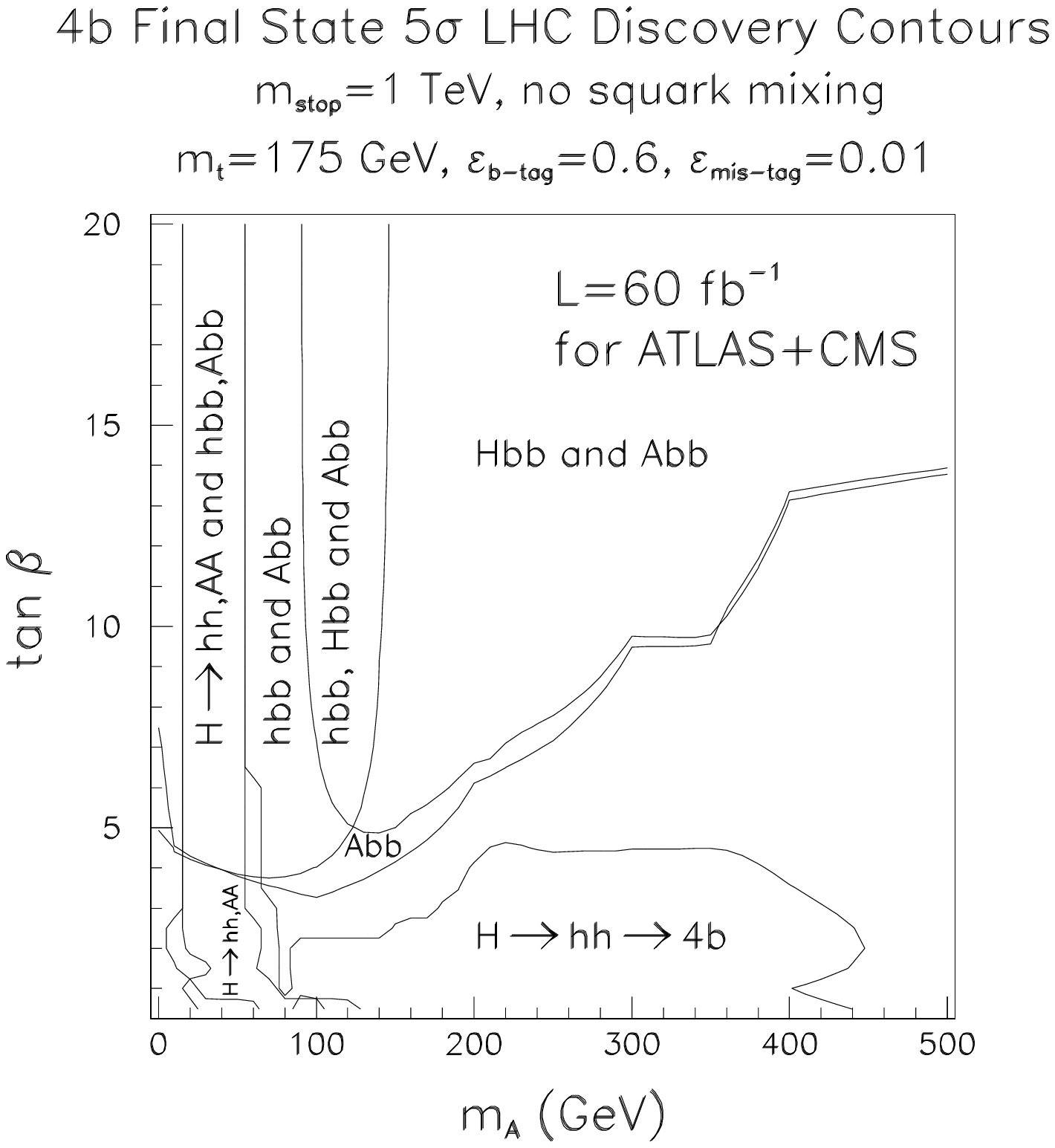,width=12.5cm}}
\smallskip
\begin{minipage}{12.5cm}       
\caption{
We show the $(\mha,\tanb)$ parameter space regions within which 
detection of (a) $gg\rta b\anti b\h\rta 4b$ ($\h=\hl,\hh,\ha$)
and/or (b) $gg\rta\hh\rta\hl\hl,\ha\ha\rta 4b$
will be possible at the $5\sigma$ level at the LHC.
We assume $L=30\fbi$ for ATLAS and CMS individually 
(combining their statistics). For (a) [(b)] we require 3 [4] tagged jets
(taking $\ebtag=0.6$, $\ectag=\ebtag/3$ and $\emistag=0.01$) with
$p_T\geq 30\gev$ [$15\gev$] and $|\eta|\leq2.5$. Radiative corrections
to Higgs masses and mixing angles \protect\cite{twoloop} are 
incorporated assuming $\mt=175\gev$, $\mstop=1\tev$ and no squark mixing.
Event rate reduction due to triggering inefficiencies is assumed to be
negligible.
}
\label{habbandhlhllhc}
\end{minipage}
\end{center}
\end{figure}

Our LHC computations for the $gg\rta b\anti b\h\rta 4b$ modes
employ 3-$b$-tagging
and incorporate precisely the same cuts and procedures
(with, in particular, approximate QCD $K$ factors included)
as outlined in Ref.~\cite{DGV3} with the following exceptions.
i) We present results for low-luminosity running ($L=30\fbi$
per detector) --- the ATLAS+CMS signal significance
is computed by combining rates for the two detectors, \ie\
we assume $L=60\fbi$ summed luminosity.
ii) Correspondingly, we employ the low-luminosity values of
$\ebtag=0.6$ and $\emistag=0.01$, requiring that 3 jets with $p_T>30\gev$
and $|\eta|\leq2.5$ be tagged as $b$-quarks.
iii) The latest (`two-loop') radiative
corrections to Higgs masses and mixing angles \cite{twoloop}
are incorporated. 
iv) We have corrected our previous failure to include identical
particle effects for the $gg\rta b\anti b b \anti b$ background
(effectively leading to a factor of $\sim 2$ too large a rate).
v) We include the $b\anti b c \anti c$ background using
$\ectag=\ebtag/3$; this results in the $b\anti b c\anti c$
background being about 60\% of the irreducible $b\anti b b\anti b$
background.
Although $b$'s could be tagged at low luminosity
for $p_T$'s as low as $15\gev$, we have found that requiring $p_T\geq30\gev$
for tagged $b$'s produces a higher statistical significance for the signal,
especially at higher masses.
4-$b$-tagging is not advantageous for the $b\anti b\h\rta 4b$
discovery modes, since one of the $b$'s produced
in association with the Higgs tends to be soft. 
Our results are presented in Fig.~\ref{habbandhlhllhc}.

In comparison to our earlier results, the region of parameter
space at large $\tanb$ for which two of the MSSM Higgs bosons 
(either the $\hl$ and $\ha$, at low $\mha$, or the $\hh$ and $\ha$,
at high $\mha$) can be observed now extends to much lower $\tanb$ values.
In Fig.~\ref{habbandhlhllhc} 
we also illustrate the complementarity of the $gg\rta b\anti b\h\rta 4b$
modes to the recently explored $gg\rta\hh\rta\hl\hl,\ha\ha\rta 4b$
discovery channel \cite{DGV4}. The plotted discovery regions for the latter 
channel are obtained
assuming low-luminosity running with $L=30\fbi$ for ATLAS
and CMS --- statistical significance is computed for ATLAS+CMS
by combining rates for the two detectors.
[For this channel, 
we require that four jets with $p_T\geq 15\gev$ and $|\eta|\leq2.5$
be tagged as $b$-quarks (using $\ebtag=0.6$ and $\emistag=0.01$).
For $\ectag\sim\ebtag/3$ and {\it four} tags, 
the $b\anti b c\anti c$ background is negligible.]
Together, the $b \anti b \h\rta 4b$ and $\hh\rta 4b$ final states 
allow discovery of one or more of the neutral MSSM Higgs bosons over a 
remarkably large portion of parameter space.
Further improvements in $b$-tagging efficiency and, especially, purity
would result in a narrowing of the inaccessible wedge
apparent in Fig.~\ref{habbandhlhllhc}
at moderate $\tanb$ which develops and widens as $\mha$ increases.

High-luminosity LHC running might or might not be advantageous
for the $gg\to b\anti b \h\to 4b$ modes. At high luminosity,
defined as $L=600\fbi$ for ATLAS+CMS,
the nominal statistical significance of the $4b$ signal increases 
by about 50\%, despite the poorer $\emistag=0.02$
mis-identification rate that enhances the dominant $b\anti b g$
background. However, as we discuss later,
triggering on the $4b$ final state becomes more problematical.
A large reduction in the triggering efficiency at high luminosity
could result in little gain in statistical significance over
low-luminosity running.

\begin{figure}[[htbp]
\vskip 1in
\let\normalsize=\captsize   
\begin{center}
\centerline{\psfig{file=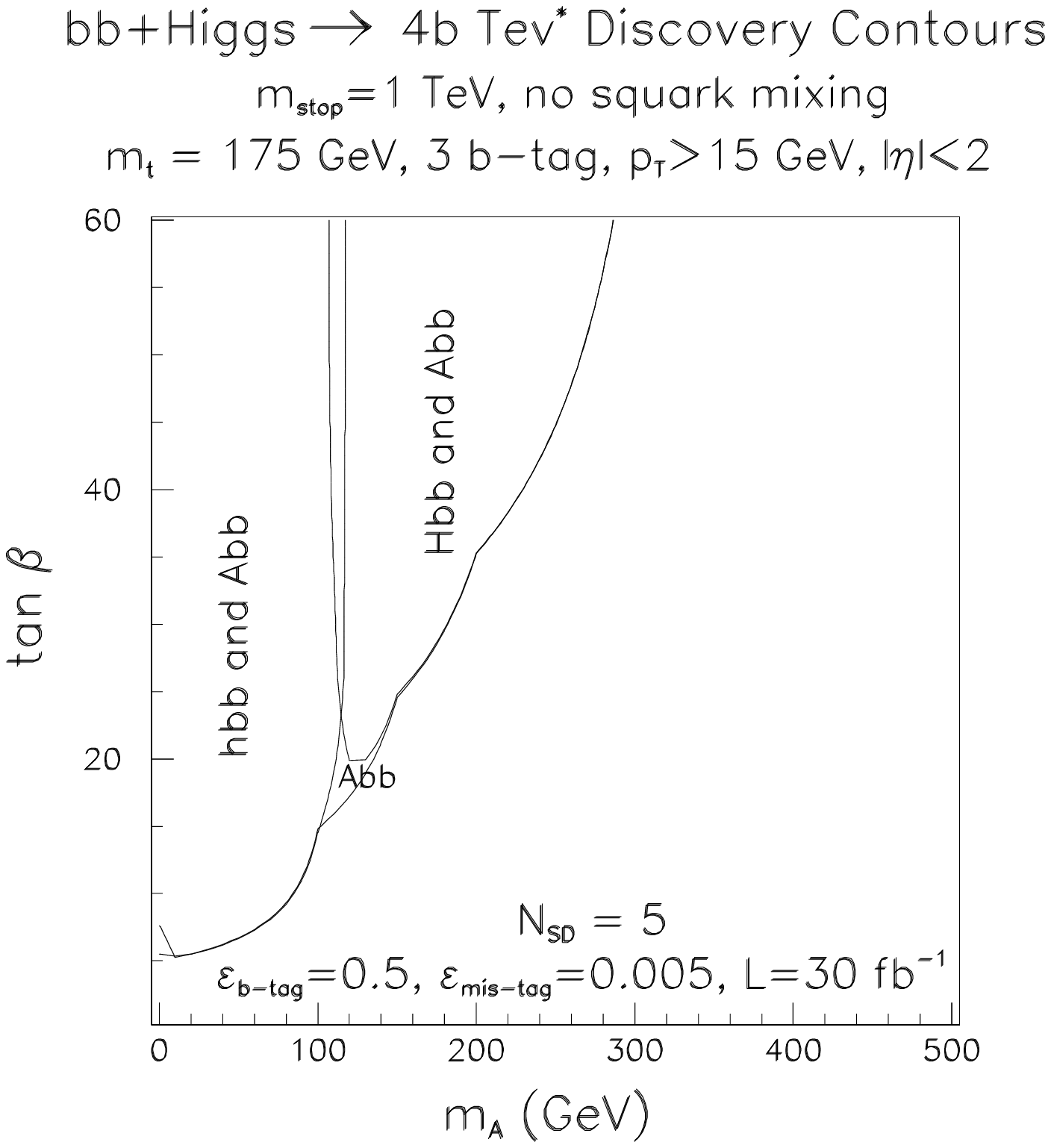,width=12.5cm}}
\smallskip
\begin{minipage}{12.5cm}       
\caption{
We show the $(\mha,\tanb)$ parameter
space regions within which $gg\rta b\anti b\h\rta 4b$ ($\h=\hl,\hh,\ha$)
can be observed at the $5\sigma$ level
at the Tevatron (operating at $\protect\sqrt s=1.8\tev$)
assuming an integrated luminosity of $L=30\fbi$.
We require 3 tagged jets (taking $\ebtag=0.5$, $\ectag=\ebtag/3$
and $\emistag=0.005$)
with $p_T\geq 15\gev$ and $|\eta|\leq2$. Radiative corrections
as in Fig.~\ref{habbandhlhllhc}.
Event rate reduction due to triggering inefficiencies is assumed to be
negligible.
}
\label{tevbbbb}
\end{minipage}
\end{center}
\end{figure}

We have also explored the parameter space regions for 
which the $gg\rta b\anti b\h\rta 4b$
channels can be detected at the Tevatron assuming that the Tev$^*$ upgrade 
achieves an integrated luminosity of $L=30\fbi$,
and that 3 jets with $p_T\geq15\gev$ and $|\eta|\leq2$
are tagged as $b$-quarks, taking $\ebtag=0.5$, $\ectag=\ebtag/3$
and $\emistag=0.005$.
The ability to achieve a very small mis-tagging
probability for light-quark and gluon jets
is crucial at the Tevatron, since the low event rates
set a premium on eliminating the large $b\anti b g$ mis-tag background.
All other procedures and cuts employed in our analysis are the
same as at the LHC, and are given in Ref.~\cite{DGV3}.
Our results for center of mass energy $\rts=1.8\tev$ 
are displayed in Fig.~\ref{tevbbbb}.
(Note the change in $\tanb$ scale relative to Fig.~\ref{habbandhlhllhc}.) 
Although MSSM Higgs discovery at the Tev$^*$ in the $gg\rta b\anti b\h\rta 4b$ 
channel is clearly limited to smaller
$\mha$ and/or much larger $\tanb$ values than at the LHC,
a significant window of opportunity is apparent. Models with 
high $\tanb\sim 60$ can be probed for $\mha$ values up to nearly $300\gev$.
We note that $gg\rta\hh\rta\hl\hl,\ha\ha\rta 4b$ allows Higgs detection
only in the $\mha\lsim 60\gev$ region (but for all $\tanb\gsim 1.5-2$)
at the Tev$^*$ \cite{DGV4}.

Decays of the neutral Higgs bosons to SUSY pair
states have been neglected in these computations. Such decays
would have small branching ratio at high $\tanb$, but, if present, could
reduce the extent of the discovery regions illustrated
at low to moderate $\tanb$.

We caution that our analysis assumes that it will be possible to
trigger on the $4b$ final states of interest with high efficiency.  
Various strategies
are currently being studied by the CDF, D0, ATLAS and CMS collaborations.
For the level-one trigger an acceptable data rate is achieved
(even for high-luminosity running at the LHC)
by simply requiring 3 jets with $p_T>30\gev$ or 
4 jets with $p_T>15\gev$ 
\cite{CMSreport}, as needed in our $b\anti b h$ and $\hl\hl$ analyses,
respectively.
The real question is whether electronic information regarding
the presence of a $b$-vertex can be fed into
level-two triggering decisions. (This will clearly be most
problematical for high-luminosity running at the LHC.)
If not, a level-two `soft'-lepton $b$-tag 
of one or more of the 4 $b$'s in the final state 
is certainly feasible; for an efficiency of $\sim 10\%$ to $\sim 15\%$ for
tagging any one of the 4 $b$'s with $p_T>15\gev$,
the net tagging efficiency for the 4-$b$ final state
would be of order 35\% to 48\%.
Full vertex information could then be recorded and incorporated much later
in the analysis. Thus, we believe that the triggering
efficiency for the events of interest will be no worse than
$\sim 35\%$, and we are hopeful that for low-luminosity running
at the LHC the electronic vertex information
could be fed in at level-two, in which
case triggering efficiency could be near 100\%. Thus, the
$4b$ final states should prove to be a very powerful
tool for Higgs detection at both the Tev$^*$ and the LHC.

\begin{center}
{\bf Acknowledgments}
\end{center}
\medskip

This work was supported in part by the U.S.~Department of Energy
under grants No.~DE-FG03-91ER40674 and No.~DE-FG03-90ER40546.
Further support was provided by the Davis Institute for High Energy Physics.


\vspace*{\fill}

\vspace*{\fill}


\bigskip

\begin{center}
{\bf References}
\end{center}
\medskip
\begin{thebibliography}{99}
\frenchspacing


\bibitem{DGV3} J.~Dai, J.~Gunion, and R.~Vega, \PLB B345 29 1995 .

\bibitem{fgianotti} F. Gianotti (ATLAS), presented
at the European Physical Society International Europhysics Conference
on High Energy Physics, Brussels, Belgium, July 27 - August 2, 1995, and
private communication. R. Kinnunen (CMS), presentation at the Tahoe CMS
Week, Tahoe City, CA, September 25-27, 1995.


\bibitem{womersley}
S. Kuhlmann, CDF note CDF/PHYS/ELECTROWEAK/PUBLIC/3342,
September, 1995. J. Womersley, private communication.


\bibitem{twoloop}
M. Carena, J.R. Espinosa, M. Quiros and C.E.M. Wagner,
CERN-TH/95-45; J.A. Casas, J.R. Espinosa, M. Quiros and A. Riotto,
\NPB B436 3 1995 ;
H. Haber, R. Hempfling and A. Hoang, CERN-TH/95-216.

\bibitem{DGV4} J.~Dai, J.~Gunion, and R.~Vega, \PLB B371 71 1996 .

\bibitem{CMSreport} Report by the CMS triggering subgroup at
CMS Week, March 1996, CERN.


\end{thebibliography}
\end{document}